\documentclass{PoS}
\newcommand{\be}{\begin{equation}}
\newcommand{\ee}{\end{equation}}
\newcommand{\bea}{\begin{eqnarray}}
\newcommand{\eea}{\end{eqnarray}}
\newcommand{\nn}{\nonumber}

\title{NP models with extended gauge groups and extra dimensions:\\ Impact on flavour observables in RS$_c$}

\ShortTitle{NP models with extended gauge groups and extra dimensions:\\ Impact on flavour observables in RS$_c$}

\author{\speaker{Fulvia De Fazio}\\
        INFN, Sezione di Bari, \\via Orabona 4, I-70126 Bari, Italy\\
        E-mail: \email{fulvia.defazio@ba.infn.it}}

\abstract{Deviations with respect to Standard Model predictions have recently shown up in angular distributions of the FCNC induced mode $B^0 \to K^{*0} \mu^+ \mu^-$. 
Within New Physics models, such tensions might be explained by
new contributions to the Wilson coefficients of the effective Hamiltonian governing this decay.
 I discuss the issue in the  framework of the Randall-Sundrum model with custodial protection (RS$_c$), giving also predictions for other rare  $B$ decays.
  }

\FullConference{The European Physical Society Conference on High Energy Physics\\
		22--29 July 2015\\
		Vienna, Austria}

\begin{document}

\section{Introduction}
Among rare $B$ decays, the mode $B \to K^* \ell^+ \ell^-$ plays a prominent role. Being  a loop-induced process within the Standard Model (SM),   possible new particles in the loops can modify the predictions for
 the numerous observables that can be measured, namely,  the branching ratio, the forward-backward lepton asymmetry,  the $K^*$ longitudinal polarization fraction in a few bins of $q^2$,  the $\ell^+ \ell^-$  invariant mass, which were measured at the B factories for $\ell=e,\,\mu$. Recently, LHCb has  found  discrepancies with respect to  SM predictions that could be hints of New Physics. Here I discuss this issue, describing the study  performed  in \cite{Biancofiore:2014wpa} within the Randall-Sundrum model \cite{RandallSundrum} with custodial protection (RS$_c$) \cite{contino}. 
I also review the results obtained    within RS$_c$
for the related modes  $B \to K^{(*)}  \nu \bar \nu$ \cite{Biancofiore:2014uba}, for which  only upper bounds on the branching ratios are available   \cite{Colangelo:1996ay,Melikhov:1998ug,Buras:2014fpa}.

\section{$B \to K^* \ell^+ \ell^-$ and $B \to K^{(*)}  \nu \bar \nu$ decays: effective Hamiltonians and general features}
  The $b \to s \ell^+ \ell^-$  transition is described by the effective Hamiltonian
\be
H^{eff}=-\,4\,{G_F \over \sqrt{2}} V_{tb} V_{ts}^* \, \Big\{C_1 O_1+C_2 O_2 
+\sum_{i=3,..,6}C_i O_i+\sum_{i=7,..,10,P,S} \left[ C_i O_i +C_i^{\prime  } O_i^{\prime  } \right] \Big\}\,\,\,. \label{hamil}
\ee
 $G_F$ is the Fermi constant and $V_{ij}$ the
elements of the Cabibbo-Kobayashi-Maskawa matrix.
Among the operators in (\ref{hamil}),  I focus here only on  $O^{(\prime)}_i$, $(i=7,\dots, 10)$:
\bea
O_7&=&{e \over 16 \pi^2} m_b ({\bar s}_{L \alpha} \sigma^{\mu \nu}
     b_{R \alpha}) F_{\mu \nu}  \hskip 0.5 cm
O_8={g_s \over 16 \pi^2} m_b \Big[{\bar s}_{L \alpha} \sigma^{\mu \nu} \Big({\lambda^a \over 2}\Big)_{\alpha \beta} b_{R \beta}\Big] \;
      G^a_{\mu \nu}  \label{mag-peng}  \,\, ,\nonumber
 \\
O_9 &=& {e^2 \over 16 \pi^2}  ({\bar s}_{L \alpha} \gamma^\mu b_{L \alpha}) \; {\bar \ell} \gamma_\mu \ell   \hskip 0.5 cm
O_{10}={e^2 \over 16 \pi^2}  ({\bar s}_{L \alpha} \gamma^\mu b_{L \alpha}) \; {\bar \ell} \gamma_\mu \gamma_5 \ell  \,\,.\label{eq-peng} \nonumber
\eea
 The corresponding primed operators are obtained reversing the quark field  chirality.
$\alpha$, $\beta$ are colour indices, $\lambda^a$ the Gell-Mann matrices, 
$F_{\mu \nu}$ and $G^a_{\mu \nu}$  denote the
electromagnetic and the gluonic field strength tensors,  $e$ and $g_s$  the
electromagnetic and the strong coupling constants, $m_b$ is the $b$ quark mass. The operators proportional to the strange quark mass  have been  neglected. Only the unprimed operators appear in  SM.

Taking into account the $K^*$ subsequent  decay into $K \pi$, the  fully differential decay width   reads:
\bea
&&\frac{d^4 \Gamma (B \to K^*[\to K \pi] \ell^+ \ell^-) }{dq^2 d\cos \theta_\ell d\cos \theta_K d\phi}=
\frac{9}{32 \pi}I(q^2,\theta_\ell, \theta_K, \phi) \,\,\, , \label{fully-diff}
\\
I(q^2,\theta_\ell, \theta_K, \phi) &=& I_1^s \sin^2 \theta_K+I_1^c \cos^2 \theta_K +(I_2^s \sin^2 \theta_K+I_2^c \cos^2 \theta_K)\cos 2 \theta_\ell +I_3 \sin^2 \theta_K \sin^2\theta_\ell \cos 2 \phi \nn \\
&+&I_4 \sin 2 \theta_K \sin 2 \theta_\ell \cos \phi +I_5 \sin 2 \theta_K \sin  \theta_\ell \cos \phi 
+(I_6^s \sin^2 \theta_K +I_6^c \cos^2 \theta_K)\cos \theta_\ell \nn  \\
&+& I_7 \sin 2 \theta_K \sin \theta_\ell \sin \phi 
+I_8 \sin 2 \theta_K \sin 2 \theta_\ell \sin \phi+I_9 \sin^2 \theta_K \sin^2 \theta_\ell \sin 2 \phi \,\label{fully-diff-1}
\eea
(the definition of the angles $\theta_K$, $\theta_\ell$ and  $\phi$ can be found in
 \cite{Faessler:2002ut,Altmannshofer:2008dz}). 
 Analogous functions $\bar I$  enter in
  the $\bar B$ meson   differential decay width 
$d^4 \bar \Gamma$, obtained replacing  in (\ref{fully-diff}) $
I_{1,2,3,4,7} \to {\bar I}_{1,2,3,4,7}$  and $
I_{5,6,8,9} \to -{\bar I}_{5,6,8,9} $  \cite{Altmannshofer:2008dz}.
 $I_i$ and $\bar I_i$  depend on the form factors parametrizing the  $B \to K^*$ hadronic matrix elements. Introducing
$S_i=\displaystyle{ \frac{I_i+\bar I_i}{ \frac{d\Gamma}{dq^2}+\frac{d\bar \Gamma}{dq^2} }}$ and  $
A_i= \displaystyle{\frac{I_i-\bar I_i}{ \frac{d\Gamma}{dq^2}+\frac{d\bar \Gamma}{dq^2}}}$, one can define the lepton forward-backward  asymmetry $A_{FB}=-\frac{3}{8} (2S_6^s+S_6^c)$,
 the longitudinal $K^*$ polarization fraction $F_L=-S_2^c$ and  
the  binned observables $<S_i>_{[q^2_1,q^2_2]}$ , with the numerators and denominators in $S_i$ separately integrated over $q^2 \in [q^2_1,q^2_2]$ .
LHCb has measured the observables
$
P^\prime_{i=4,5,6,8}=\displaystyle{\frac{S_{i=4,5,7,8}}{\sqrt{F_L(1-F_L)}}}
$  \cite{DescotesGenon:2012zf},  finding that the measurement of $P^\prime_5$, performed in  bins of $q^2$, deviates from SM predictions for  low $q^2$ values \cite{Aaij:2013iag,LHCbnew}. 

For  $B \to K^{(*)}  \nu \bar \nu$ decays,
the most general $b \to s \nu \bar \nu$ effective Hamiltonian is: $
H_{eff}=C_L O_L+ C_R O_R $,
where  $O_{L,R}=({\bar b} s)_{V \mp A}({\bar \nu} \nu)_{V-A}$  \cite{Buras:1998raa}.
In  SM the contribution of $O_R$ is negligible and $C_L^{SM}= {G_F \over \sqrt{2}} {\alpha \over 2 \pi \sin^2\theta_W} V_{tb}^* V_{ts} X(x_t)$.  $\alpha$ is the fine structure constant  at the $Z^0$ scale and $\theta_W$  the Weinberg angle.  
The function $X$ depends on the ratio of the top and  $W$ masses $x_t=m_t^2/M_W^2$ \cite{Inami:1980fz}.
In NP scenarios, also $O_R$ can be present, and $C_{L,R}$ assume   model specific  values.
It is useful to introduce two parameters,  
$\epsilon^2=\displaystyle{\frac{|C_L|^2+|C_R|^2}{|C_L^{SM}|^2} }$ and $ \eta=-\displaystyle{\frac{{\rm Re}\left( C_L C_R^* \right)}{|C_L|^2+|C_R|^2}}$,
sensitive to deviations from  SM where $\left( \epsilon,\, \eta \right)_{SM}=(1,0)$ \cite{Melikhov:1998ug} .
$\eta$ probes the presence of  $O_R$, while $\epsilon$  measures the deviation of  $C_L$ from its SM value. 
Predictions   in NP extensions can be expressed in terms of $\eta$ and $\epsilon$. 
In \cite{Biancofiore:2014uba} the branching fractions and  the spectra in the normalized  neutrino pair invariant mass  $s_B=q^2/m_B^2$ have been computed and, for the decay  $B \to K^* \nu \bar \nu$, also  the polarization fractions for longitudinally and transversely polarized $K^*$: $F_{L,T}=\frac{1}{\Gamma} \, \int_0^{1-{\tilde m}_{K^*}^2} ds_B \, \displaystyle{\frac{d \Gamma_{L,T} }{ds_B}}$. 
Denoting the branching ratio for a transversely polarized $K^*$ as ${\cal B}_T={\cal B}(B \to K^*_{h=-1} \,\nu \bar \nu) +{\cal B}(B \to K^*_{h=+1} \,\nu \bar \nu) $, 
other  observables are 
$
R_{K/K^*}=\displaystyle{\frac{{\cal B}(B \to K \,\nu \bar \nu)  }{{\cal B}_T}}$, sensitive to  $\eta$, and $ A_T=\displaystyle{\frac{{\cal B}(B \to K^*_{h=-1} \, \nu \bar \nu) -{\cal B}(B \to K^*_{h=+1} \, \nu \bar \nu) }{{\cal B}_T}} $,
 expected to  be affected by   a small  hadronic uncertainty  \cite{Melikhov:1998ug}.

\section{Randall-Sundrum model with custodial protection}
The RS model is defined in  a five-dimensional   spacetime  with 
metric 
$
ds^2=e^{-2 k y} \eta_{\mu \nu} dx^\mu dx^\nu - dy^2$, where 
$\eta_{\mu \nu}=diag(+1,-1,-1,-1)$, $x$  denote the ordinary 4D  coordinates and 
 $y$ varies in the range $0 \le y \le L$  ($y=0$ is called UV brane,   $y=L$  IR brane).
The parameter $k$  is  fixed to  $k=10^{19}$ GeV to adress the hierarchy problem through a geometrical mechanism.  
The custodially protected  variant of  the model   is based on the   group 
$ SU(3)_c \times SU(2)_L \times SU(2)_R \times U(1)_X \times P_{L,R}$ \cite{contino}.
The  discrete  $P_{L,R}$ symmetry  implies a mirror action of the two $SU(2)_{L,R}$ groups,   preventing
 large $Z$ couplings to left-handed fermions.
The group  is broken to the  SM  group by   boundary conditions (BC) on the UV brane; moreover, Higgs-driven spontaneous symmetry breaking occurs, as in  SM.
All  fields can propagate in the bulk, except for the Higgs localized close to the IR brane.

Due to the compactification of $y$,   towers of Kaluza-Klein (KK) excitations exist for all particles. The zero modes are identified with  SM particles.
To distinguish particles having a SM correspondent from those without it,  Neumann BC  on both branes (++) are imposed, while  Dirichlet BC on the UV brane and Neumann BC on the IR one (-+) are chosen for fields without  SM partners.

The enlarged gauge group leads to new gauge bosons. In the case of $SU(2)_L$ and $SU(2)_R$ they are $W_L^{a,\mu}$  and $W_R^{a, \mu}$  ($a=1,2,3$),  respectively, while the 
 $U(1)_X$ gauge field  is $X_\mu$.
Charged gauge bosons are defined as $
W_{L(R)\mu}^\pm=\frac{W^1_{L(R)\mu} \mp i W^2_{L(R)\mu}}{\sqrt{2}}$. As for neutral fields, $W_R^3$ and $X$  mix to give $Z_{X }$ and $B$; $B$ mixes with
$W_{L \,}^3$ giving $Z$ and $A$ fields. 
 Zero modes and higher KK modes of gauge fields also mix. Neglecting modes with KK number larger than $1$, mixings occur, $ (W_L^{\pm (0))}\,\, W_L^{\pm (1)} \,\, W_R^{\pm (1)}) \to     (W^\pm \,\,  W_H^\pm \, \, W^{\prime\,\, \pm})$ and  $ ( Z^{(0)} \,\, Z^{(1)} \,\, Z_X^{(1)}) \to     (Z \,\,  Z_H \, \, Z^\prime)$ \cite{Albrecht:2009xr}.

In the Higgs sector, the Higgs field 
$H(x,y)$  transforms as a bidoublet under $SU(2)_L \times SU(2)_R$ and as a singlet under $U(1)_X$. 
It contains two charged and  two neutral components. Only one of the two  neutral fields,  $h^0$, has  a non-vanishing  vacuum expectation value   $v=246.22$ GeV, as in  SM. 

Moving to fermions, SM left-handed doublets fit in  bidoublets of $SU(2)_L \times SU(2)_R$, together with two new fermions.
 Right-handed up-type quarks are singlets;    neutrinos are only left-handed.
  Right-handed down-type quarks and  charged leptons   transform as $(3,1) \oplus (1,3)$   $SU(2)_L \times SU(2)_R$ multipltes in which   additional new fermions are also present.
   The relation $Q=T^3_L+T^3_R+Q_X$ holds among 
the electric charge $Q$,  the third component of the $SU(2)_L$ and $SU(2)_R$ isospins $T_{L,R}^3$ and the charge $Q_X$ .
 The profiles of zero-mode fermions involve the fermion bulk mass, which is  the same for fermions in the same $SU(2)_L \times SU(2)_R$ multiplet. 

As  in  SM, quark flavour eigenstates undergo a rotation to give mass eigenstates. Denoting by ${\cal U}_{L(R)}$, ${\cal D}_{L(R)}$ the rotation matrices of up-type left (right) and down-type left (right) quarks, respectively,  the CKM matrix is   $V_{CKM}={\cal U}_L^\dagger {\cal D}_L$.
Their matrix elements are involved in the Feynman rules of tree-level flavour-changing neutral currents that exist in the model, mediated by
$Z,\,Z^\prime, \, Z_H$,  and  by the first KK mode of the photon and of the gluon. 
Such elements
 depend on the 5D Yukawa couplings  $\lambda_{ij}^{u,d}$  of up and down-type quarks,  constrained   to reproduce quark masses and CKM elements.
Adopting   the  assumption of real and symmetric $\lambda^{u,d}$ matrices, one is left with six independent entries among their elements, namely $
\lambda^u_{12} \,\,\, ,   \lambda^u_{13} \,\,\, , \lambda^u_{23} \,\,\, ,
\lambda^d_{12} \,\,\, ,  \lambda^d_{13} \,\,\, , \lambda^d_{23}  \,\,\,, $ which, together with the bulk mass parameters,  represent  the   set of  numerical inputs of our study.

\section{$B \to K^* \ell^+ \ell^-$ and  $B \to K^{(*)} \nu \bar \nu$ decays in RS$_c$ }
In the RS$_c$ model the Wilson coefficients,  $C^{RS}=C^{SM}+\Delta C$, have been derived in \cite{buras2},
 except for    $C_7$ and $C_7^\prime$
 computed in \cite{Biancofiore:2014wpa}  with the same assumptions adopted in  \cite{buras2} . Different computational schemes for $C_7^{(\prime)}$ were used in \cite{Blanke:2012tv}. 

The new  contributions $\Delta C$ are obtained  scanning  the  parameter space.
In  \cite{Biancofiore:2014wpa,Biancofiore:2014uba} the quark bulk mass parameters and the independent entries  of the matrices  $\lambda^{u,d}$ 
have been fixed imposing  quark masses and CKM constraints, as well as  constraints  derived in \cite{neubertRS} using  the  measurements of  the coupling $Z{\bar b}b$, of the $b$-quark left-right asymmetry parameter and of the forward-backward asymmetry for $b$ quarks.
The parameter space is further reduced   imposing that   
${\cal B}(B \to K^* \mu^+ \mu^-)$ and ${\cal B}(B \to X_s \gamma)$  lie within the 
$2\sigma$ range of  the measurements \cite{Lees:2012tva,Amhis:2012bh}. 
For further  details  I refer to \cite{Biancofiore:2014wpa} .
 %%%%
\begin{figure}[h]
\begin{center}
\includegraphics[width = 0.38\textwidth]{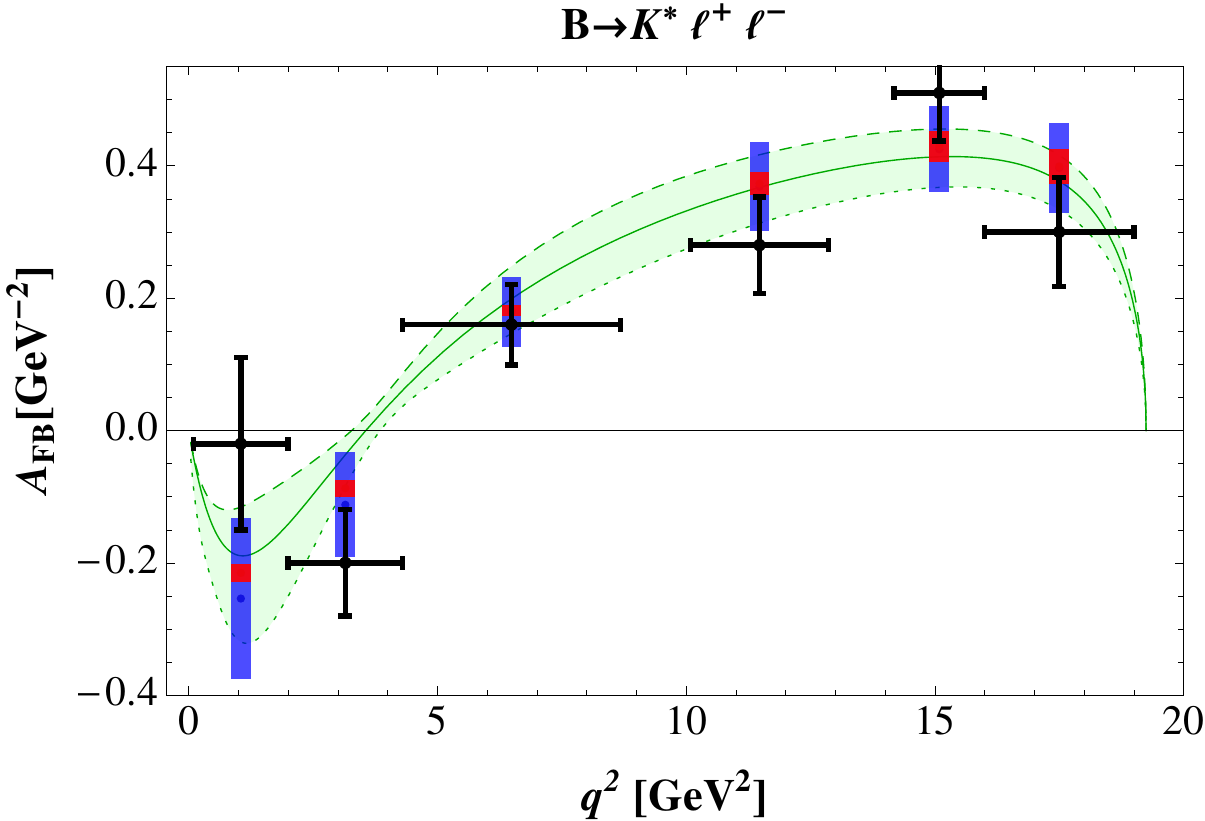}\hskip 0.5 cm  \includegraphics[width = 0.38\textwidth]{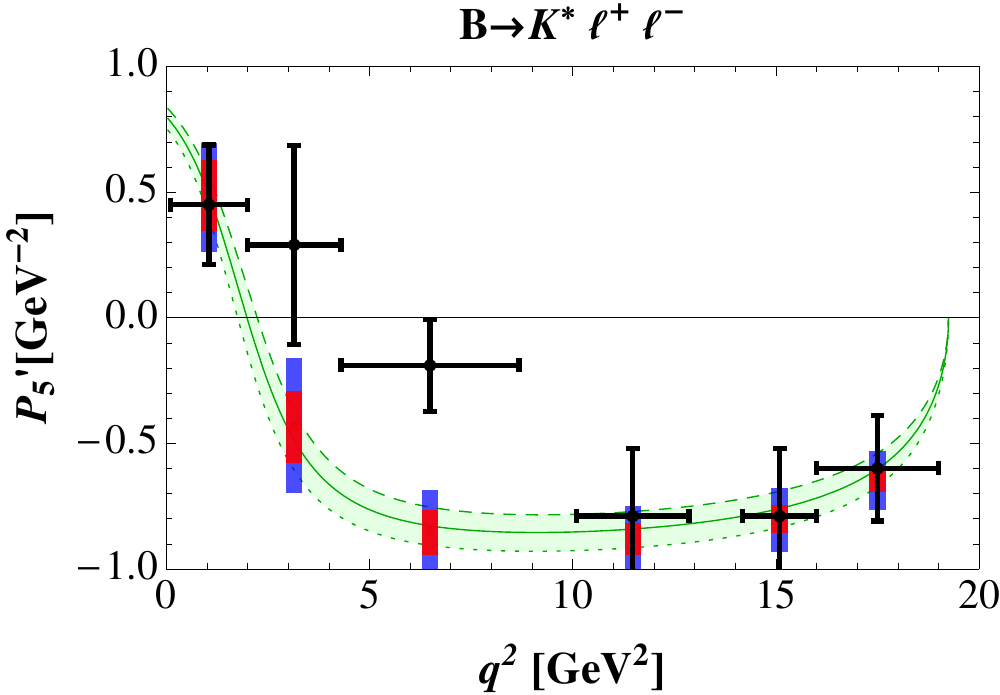}
\end{center}
\caption{$A_{FB} $ (left) and  $P^\prime_5$ (right) in  $B^0\to K^{*0} \mu^+ \mu^-$. The green band is  the SM result, including the uncertainty of  the form factors. The red and blue vertical bars are the RS$_c$ result, 
without or with  the uncertainty in form factors. The black dots, with their error bars, are the LHCb measurements in  \cite{Aaij:2013iag}.}\label{fig:AFB}
\end{figure}
%%%%%

In Fig. \ref{fig:AFB}  SM and RS$_c$ predictions for $A_{FB}$ and $P^\prime_5$ are compared,  varying the model parameters and including the uncertainty on  the  form factors  computed   in \cite{Ball:2004rg} using  light-cone QCD sum rules \cite{Colangelo:2000dp}.
The   form factor uncertainty has an impact on the SM results, except for the position of the zero in $A_{FB}(q^2)$, almost free of uncertainty.
In RS$_c$, deviations from SM are small, and discrepancy with data is found  as well. For  $B \to K^* \tau^+ \tau^-$  no data are available at present  \cite{Biancofiore:2014wpa} .

\begin{figure}[t!]
\begin{center}
\includegraphics[width = 0.4\textwidth]{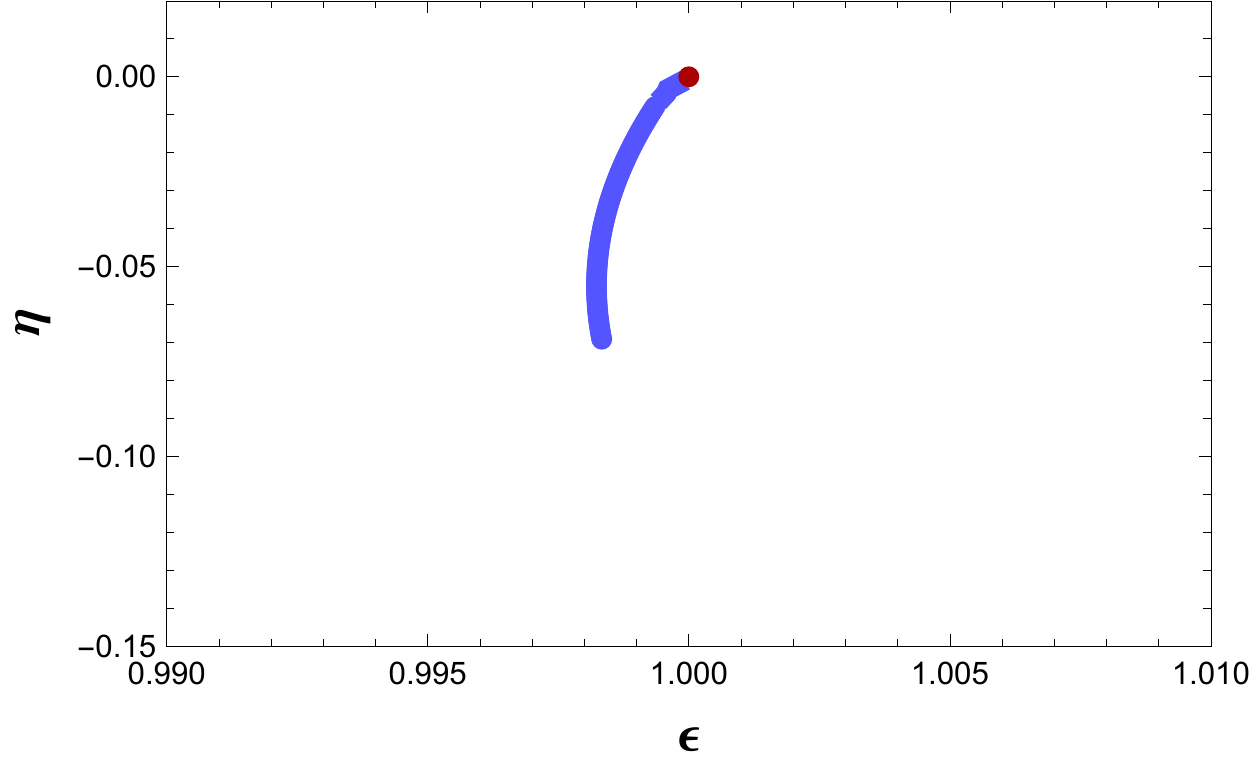}\hskip 0.5 cm  \includegraphics[width = 0.38\textwidth]{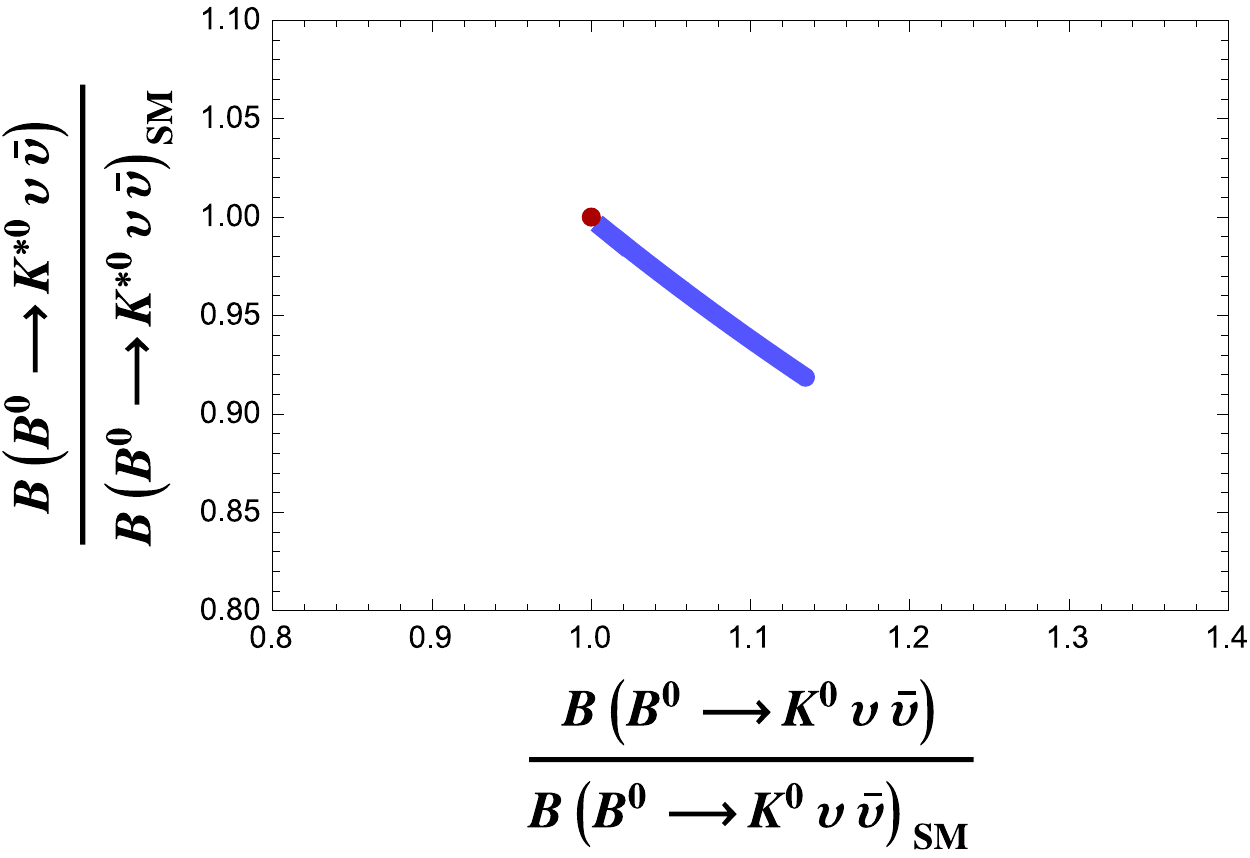}
\caption{Left: Correlation between the parameters $\eta$ and  $\epsilon$ in the RS$_c$ model (blue curve). The red dot corresponds to SM.
Right: Correlation between ${\cal B}(B^0 \to K^0 \nu \bar \nu)$  and  ${\cal B}(B^0 \to K^{*0} \nu \bar \nu)$ (blue curve) normalized to the corresponding SM values (red dot) obtained for  the central value of the form factors.}
\label{fig:etaeps}
\end{center}
\end{figure}

Considering the modes $B \to K^{(*)} \nu \bar \nu$, in Fig. \ref{fig:etaeps} the $(\epsilon,\,\eta)$ correlation plot shows  that the largest value of $\eta$ in RS$_c$ is $\eta=-0.075$,  compared to  the SM value $\eta=0$. This is the consequence of the non vanishing role of the operator $O_R$  in the model.
The branching ratios in RS$_c$, 
$
{\cal B}(B^0 \to K^0 \nu \bar \nu)_{RS}\in [3.45 - 6.65]\, \times 10^{-6}$  and 
$ {\cal B}(B^0 \to K^{*0} \nu \bar \nu)_{RS} \in [6.1 - 14.3] \times 10^{-6}$,
span a range larger than in SM, $
{\cal B}(B^0 \to K^0 \nu \bar \nu)_{SM}=(4.6 \pm1.1)\, \times 10^{-6}$ $, 
{\cal B}(B^0 \to K^{*0} \nu \bar \nu)_{SM}=\,(10.0 \pm2.7) \times 10^{-6} $.
The right panel of Fig. \ref{fig:etaeps} displays the results  in correspondence to  the central value of the form factors, showing 
an anticorrelation between the branching ratios of the two modes
 in RS$_c$.

The  pattern of correlations among the various observables is interesting \cite{Biancofiore:2014uba}.
 ${\cal B}(B^0 \to K^{*0} \nu \bar \nu)$ and $F_L$ are correlated, while ${\cal B}(B^0 \to K^{*0} \nu \bar \nu)$ and $A_T$ are anticorrelated, as well as $R_{K/K^*}$ and $F_L$, a pattern that  can be viewed as a specific feature of  RS$_c$.
Similar features appear in   the   decays  $B_s \to (\phi,\,\eta,\, \eta^\prime \, f_0(980)) \nu \bar \nu$  \cite{Biancofiore:2014uba}.

\section{Conclusions}

 In the RS$_c$ model, deviations with respect to SM predictions are found  in several observables relative to the modes $B \to K^* \ell^+ \ell^-$ and $B \to K^{(*)} \nu \bar \nu$ , even though  small. Correlations among observables exist, that  can be used    to discriminate this model from other  NP scenarios. 

{\bf Acknowledgments.} I thank P. Biancofiore, P. Colangelo and E. Scrimieri for collaboration on the topics discussed here,  and A. J. Buras for useful discussions.

\end{document}